\documentclass[11pt]{article}
\usepackage{fullpage,times,amsmath,amssymb,amsthm,hyperref,graphicx,scalerel}
\usepackage[T1]{fontenc}
 
\newtheorem{theorem}{Theorem}[section]
\newtheorem{lemma}[theorem]{Lemma}
\newtheorem{corollary}[theorem]{Corollary}
\newtheorem{problem}[theorem]{Problem}
\newtheorem{observation}[theorem]{Observation}
\newtheorem{hypothesis}[theorem]{Hypothesis}
 
\newcommand{\eps}{\varepsilon}
\newcommand{\OO}{\widetilde{O}}

\newcommand{\down}[1]{\left\lfloor #1\right\rfloor}

\newcommand{\F}{\mathbb{F}}
\newcommand{\AAA}{\widehat{A}}
\newcommand{\BBB}{\widehat{B}}
\newcommand{\BBBB}{\mathcal{B}}

\newcommand{\Indicator}[1]{\mathbf{1}[#1]}

\newcommand{\bits}{\F_2^n}
\DeclareMathOperator\supp{\mathrm{supp}}

\newcommand\reallywidehat[1]{\arraycolsep=0pt\relax%
\begin{array}{c}
\stretchto{
  \scaleto{
    \scalerel*[\widthof{\ensuremath{#1}}]{\kern-.5pt\bigwedge\kern-.5pt}
    {\rule[-\textheight/2]{1ex}{\textheight}} 
  }{\textheight} %
}{0.5ex}\\           
#1\\                 
\rule{-1ex}{0ex}
\end{array}
}

\begin{document}

\title{The 3-Uniform Hyperclique Hypothesis is False}

\author{Timothy M. Chan\thanks{Siebel School of Computing and Data Science, University of Illinois at Urbana--Champaign, USA (\{tmc, itakura3, 
zm25, yyu51\}@illinois.edu)}\and
Reina Itakura$^\ast$ \and
Zhiyuan Ma$^\ast$ \and
Yuancheng Yu$^\ast$
}

\maketitle

\begin{abstract}
We present truly subquartic algorithms for detecting or counting 4-hypercliques in a 3-uniform hypergraph with $n$ vertices, disproving a popular hypothesis in fine-grained complexity.
The result immediately implies a faster algorithm for finding a $k$-hyperclique in a 3-uniform hypergraph for all constant $k\ge 4$, and a faster exponential-time algorithm for Max-3-SAT.
\end{abstract}

\section{Introduction}

Recently, with the advances of AI tools, we have seen the refutations of a number of well-known hypotheses in fine-grained complexity: the 3SUM, APSP, and Exact-Triangle Hypotheses~\cite{AlmanV26}, as well as SETH and ETH~\cite{OpenAI26,Antropic26}.  In this paper, we add one more to the list\ldots 

In the \emph{$h$-uniform $k$-hyperclique} problem ($k>h$), we are 
given an $h$-uniform hypergraph with $n$ vertices and want to detect the
existence of a $k$-hyperclique (i.e., a size-$k$ set such that any
size-$h$ subset is a hyperedge).
The $h=2$ case corresponds to the standard $k$-clique problem in graphs, for which there
exist algorithms running in better than $n^k$ time by a straightforward reduction to matrix multiplication
(the current fastest algorithm runs in $O(n^{\omega k/3})$ time for $k$ divisible by 3,
where $\omega$ is the matrix multiplication exponent).
However, for the $h=3$ case, it is not clear if matrix multiplication machineries help.
The current best algorithms for $h\ge 3$ achieved only polylogarithmic factor speedups~\cite{Nagle10,AbboudFS26}.
Lincoln, Vassilevska Williams, and Williams~\cite{LincolnWW18}
introduced the following hypothesis:

\begin{hypothesis}[$h$-Uniform $k$-Hyperclique Hypothesis]
Let $k>h>2$.
Finding a $k$-hyperclique in an $h$-uniform hypergraph on $n$ vertices
requires $n^{k-o(1)}$ time.
\end{hypothesis}

Lincoln, Vassilevska Williams, and Williams provided some reasons why 
the hypothesis is believable. 
For $h\ge 3$, they wrote:
``the fastest algorithms for the $\ell$-hyperclique problem run in $n^{\ell-o(1)}$ time, and this is not for
lack of trying. Many researchers \ldots have attempted to
design a faster algorithm, for instance by mimicking the matrix multiplication approach for $k$-Clique. However, in doing
this, one needs to design a nontrivial algorithm for a generalized version of matrix multiplication.'' 
They went on to show that the border rank of this generalized
``$(k-1)$-wise matrix product'' problem is $n^k$, and argued that this rules out ``arithmetic circuit'' approaches.  (However, Brand, Kaski, and Wang \cite{BrandKW26} noted inadequacy to this rank argument, and proposed better notions of rank, though without providing 
new lower bounds for hyperclique.)

As a further argument,
Lincoln, Vassilevska Williams, and Williams also recalled known reductions, showing that a faster algorithm for
$h$-uniform $k$-hyperclique would imply faster algorithms for
other problems such as Max-$h$-SAT (and other constrained satisfaction problems)
for $h\ge 3$~\cite{Wil07},
and Max Exact $k$-Clique for $h\ge 4$
(although the Max Exact $k$-Clique Hypothesis was recently refuted along with Exact Triangle~\cite{AlmanV26}).

Over the last few years,
the Hyperclique Hypothesis has  gained prominence as a means to proving
conditional lower bounds
for many problems, and there is now a long list of applications:
detecting induced 4-cycles in graphs~\cite{DalirrooyfardW22},
detecting other small subgraphs in graphs~\cite{BringmannS21,FischerKRS25},
finding hypercycles in hypergraphs~\cite{DBLP:conf/icalp/FuLR25},
LCAs in DAGs~\cite{MathialaganWX22},
problems in logic and database theory~\cite{BringmannFK19,AnGIJKN22},
constraint satisfaction problems~\cite{KunnemannM20},
the $d$-dimensional maximum subarray problem~\cite{WilliamsX20},
diameter in geometric intersection
graphs~\cite{BringmannKKNP22,ChanCGKLZ26a,ChanCGKLZ26}, Klee's measure problem~\cite{Kunnemann22,GorbachevK23},
and other geometric problems~\cite{Chan23,ChanHY23,AngrickBGK26}.
The majority of these reductions use the $h=3$ case.  The 3D Klee's measure
problem is a good example: an easy near-$n^{3/2}$ lower bound~\cite{Chan10} was known
for combinatorial algorithms under the Combinatorial $k$-Clique Hypothesis,
but faster algorithms might still conceivably exist using fast matrix multiplication.  K\"unnemann~\cite{Kunnemann22} conditionally ruled out this possibility by establishing a tight near-$n^{3/2}$
lower bound for non-combinatorial algorithms under the 3-Uniform Hyperclique Hypothesis.

Our main result is an $O(n^{4-\eps})$-time (deterministic) algorithm for 
finding a 4-hyperclique in a 3-uniform hypergraph, for some constant $\eps>0$.
Immediately, by grouping vertices into clusters of size $\down{k/4}$,
we obtain an $O(n^{k-\eps\down{k/4}})$-time algorithm for $k$-hyperclique in a 
3-uniform hypergraph for all $k\ge 4$.  This thus refutes the 3-Uniform Hyperclique
Hypothesis.  It also immediately implies an $O(2^{(1-\eps/4+o(1))n})$-time
algorithm for Max-3-SAT~\cite{Wil07,LincolnWW18}, which is a significant improvement over  the previous
best time bound of $O(2^{n-n/O(\log^2n\log^2\log n)})$ by Alman, Chan, and Williams~\cite{AlmanCW16} (and disproving what some have called the Max-3-SAT Hypothesis~\cite{LincolnWW18}).

We have actually found two  different proofs of the result (both using AI\@)!   Both are relatively short.
Both approaches do not require inventing new techniques to solve
a generalized 3-wise matrix product problem,  but instead just reduce to standard rectangular matrix multiplication.
The first (described in Section~\ref{sec:first}) gives a better exponent; it first reduces the problem to some
color-alternating 4-cycle counting problem, and uses known anti-commuting families of matrices to cancel out 4-cycles that do not satisfy the color-alternation constraint.
The second (described in Section~\ref{sec:second}) is  shorter, and solves the problem directly using formulas that may appear ``magical'', involving univariate polynomials and the Walsh--Hadamard transform, working in the Hamming cube.
The reader may choose to read either
Section~\ref{sec:first} or~\ref{sec:second}, or both in any order.

The main takeaway is that clever (but relatively simple) combinations of algebraic techniques can achieve very surprising things,
more so than previously thought.  Such combinations may be difficult to find by (and have eluded) human researchers, but may be easier to find now in the age of AI.

\paragraph{Notation.}
Let $[n]=\{0,\ldots,n-1\}$.
For a set $S$, let $2^S$ denote the set of all subsets of $S$, and
${S\choose k}$ denote the set of all subsets of $S$ of size $k$,
and ${S\choose \le k}$ denote the set of all subsets of $S$ of size at most $k$.
Let ${n\choose \le k} = {n\choose 0}+\cdots+{n\choose k}$.
Let $\Indicator{E}$ be the indicator function for the predicate $E$.
We use $\lg$ for base-2 logarithms when the base matters.
Let $M(n_1,n_2,n_3)$ denote the complexity of multiplying an $n_1\times n_2$
and an $n_2\times n_3$ matrix.

\section{First Approach}\label{sec:first}

\subsection{Reducing 3-Uniform 4-Hyperclique to An Alternating 4-Cycle Counting Problem}

We will solve the following more general form of the 3-uniform 4-hyperclique problem
(in the case of 0-1 entries, this allows us to not just detect but count the number of 4-hypercliques, and we can also report one 4-hyperclique, if exists, by binary searching with an extra logarithmic factor):

\begin{problem}[3-Uniform 4-Hyperclique Generalized Counting]\label{prob:4hyperclique}
Given four 3-dimensional $n\times n\times n$ arrays $P,Q,R,S$,
compute
\[ 
 \sum_{a,b,c,d\in[n]} P[a,b,c]Q[a,b,d]R[a,c,d]S[b,c,d].
\] 
\end{problem}

Our first step is to reduce to a different problem:

\begin{problem}[Alternating 4-Cycle Generalized Counting] \label{prob:4matrix}
Given $n_A$ symmetric $n\times n$ matrices $A_0,\ldots,A_{n_A-1}$
and $n_B$ symmetric $n\times n$ matrices $B_0,\ldots,B_{n_B-1}$, compute 
\[ \sum_{x,y,z,z'\in [n],\ a\in[n_A],\ b\in [n_B]} A_a[x,z]B_b[z,y]A_a[y,z']B_b[z',x].
\]
\end{problem}

This can be viewed as a generalized counting problem about 4-cycles in colored multigraphs: in the 0-1 case,
we are given a multigraph with $n$ vertices where each type-A edge has a color in $[n_A]$, 
 each type-B edge has a color in $[n_B]$, and we want to count 4-cycles whose edges
alternate between type A and type B, such that the colors of opposite edges are identical
(i.e., of the form $a,b,a,b$).

\begin{lemma}
If Problem~\ref{prob:4matrix} can be solved in $T(n,n_A,n_B)$ time, then
Problem~\ref{prob:4hyperclique} can be solved in $O(T(n^{1+o(1)},n^{1+o(1)},n^{1+o(1)}))$ time.
\end{lemma}
\begin{proof}
By Behrend's construction~\cite{Behrend46} (which has been
used before in graph algorithms and fine-grained reductions, e.g., \cite{DBLP:journals/rsa/Alon02,AbboudBHS22,JinX23}), there exists a set $S\subseteq [M]$ with no 3-term arithmetic progressions, such that $|S|\ge M^{1-o(1)}$.  Such a set can be constructed in $\OO(M)$ time. 
Let $T=\{(x,x+d,x+2d): x\in [M],\ d\in S\}$. 
Define
\[\begin{array}{llll}
L_{12}(x,y)&=&(x,y,y+(y-x)) &\mbox{if it is in $T$, and undefined otherwise;}\\
L_{23}(y,z)&=&(y-(z-y),y,z) & \mbox{if it is in $T$, and undefined otherwise;}\\
L_{13}(x,z)&=&(x,(x+z)/2,z) & \mbox{if it is in $T$, and undefined otherwise.}
\end{array}\]
Observe that
\begin{equation}\label{eqn:determined}
\mbox{If $L_{12}(x,y)$, $L_{23}(y,z)$, and $L_{13}(x,z)$ are all defined,
then $L_{12}(x,y)=L_{23}(y,z)=L_{13}(x,z)$.}
\end{equation}
This is because if $y-x,z-y,(z-x)/2\in S$, then they must all be equal since
$S$ does not contain a 3-term arithmetic progression.

Since $|T|\ge M^{2-o(1)}$, we can ensure $|T|\ge n$ by setting $M=n^{1/2+o(1)}$.
Given $P,Q,R,S$,
define the following matrices:
for $x_a,y_a,z_a,\ldots, x_d,y_d,z_d\in [3M]$,
\begin{eqnarray*}
A_{(y_a,x_c)}[(y_b,y_c,1),(x_a,x_b,2)] &=& P[L_{12}(x_a,y_a),L_{12}(x_b,y_b),L_{12}(x_c,y_c)]\\
B_{(z_b,y_d)}[(x_a,x_b,2),(z_a,x_d,3)] &=& Q[L_{13}(x_a,z_a),L_{13}(x_b,z_b),L_{12}(x_d,y_d)]\\
A_{(y_a,x_c)}[(z_a,x_d,3),(z_c,z_d,4)] &=& R[L_{23}(y_a,z_a),L_{13}(x_c,z_c),L_{23}(x_d,z_d)]\\
B_{(z_b,y_d)}[(z_c,z_d,4),(y_b,y_c,1)] &=& S[L_{23}(y_b,z_b),L_{23}(y_c,z_c),L_{23}(y_d,z_d)].
\end{eqnarray*}
We can set all other entries to 0, and make the matrices symmetric.
There are $O(M^2)$ matrices of dimension $O(M^2)\times O(M^2)$,
with $M^2=n^{1+o(1)}$.

By (\ref{eqn:determined}), it is straightforward to verify that
\begin{eqnarray*}
\lefteqn{ 4\sum_{a,b,c,d\in T} P[a,b,c]Q[a,b,d]R[a,c,d]S[b,c,d] } \\
&=& 
\sum_{\scriptsize\begin{array}{c}(y_b,y_c),(x_a,x_b),(z_a,x_d),(z_c,z_d),\\(y_a,x_c),(z_b,y_d)\in [3M]^2\end{array}}
\begin{array}{l}
A_{(y_a,x_c)}[(y_b,y_c,1),(x_a,x_b,2)]\cdot B_{(z_b,y_d)}[(x_a,x_b,2),(z_a,x_d,3)]\cdot \\
A_{(y_a,x_c)}[(z_a,x_d,3),(z_c,z_d,4)]\cdot B_{(z_b,y_d)}[(z_c,z_d,4),(y_b,y_c,1)].
\end{array}
\end{eqnarray*}
(The reason for the factor 4 is that each alternating 4-cycle is counted 4 times, where the last coordinates may follow the sequence $\langle 1,2,3,4\rangle$,
$\langle 2,1,4,3\rangle$, $\langle 3,4,1,2\rangle$, and $\langle 4,3,2,1\rangle$.)
\end{proof}

\subsection{Solving the Alternating 4-Cycle Counting Problem}

To solve Problem~\ref{prob:4matrix},
a naive attempt is to compute an  intermediate matrix
\[ P[x,y]\ :=\ \sum_{z\in[n],\ a\in[n_A],\ b\in[n_B]}
A_a[x,z]B_b[z,y],
\]
which looks similar to matrix multiplication; we then compute the sum 
\[ \sum_{x,y\in[n]} P[x,y]P[y,x].
\]
Unfortunately, this sum may count 4-cycles that don't necessarily have matching
opposite colors (colors $a,b,a',b'$ where we cannot guarantee that $a=a'$ and $b=b'$).  

The key ingredient to fix this is the following algebraic lemma.  The lemma lets  us ``cancel'' out 4-cycles that don't have matching opposite colors, 
by modifying the above formulas with cleverly chosen coefficients.
The coefficients need to satisfy a certain ``anti-commuting'' property: when summing over some number of indices $\ell$,
we want the contributions of $h^{(\ell)}_{a,b}h^{(\ell)}_{a',b'}$ to equal the contributions of $-h^{(\ell)}_{a,b'}h^{(\ell)}_{a',b}$,
unless $a=a'$ and $b=b'$.  (Indeed, the proof of the lemma, which we will provide in the
next subsection, makes use of known anti-commuting families of matrices.)

\begin{lemma}[Main algebraic lemma]\label{lem:algebraic}
Let $n_A=2^{m/2}$, 
$n_B={m\choose k}$, and $L=n_A\cdot {m\choose \le k/2}$,
where $m$ is divisible by $8$ and $k$ is even.
There exist numbers $h_{a,b}^{(\ell)}\in\{-1,0,1\}$
and $w^{(\ell)}\in\mathbb{Q}$, for each $a\in [n_A]$, $b\in [n_B]$, and
$\ell\in [L]$, such that for all $a,a'\in[n_A]$ and $b,b'\in[n_B]$,
\begin{equation}\label{eqn:main}
\frac{1}{2}\sum_{\ell\in[L]}w^{(\ell)}(h_{a,b}^{(\ell)}h_{a',b'}^{(\ell)}
+ h_{a,b'}^{(\ell)}h_{a',b}^{(\ell)})\ =\ \Indicator{a=a'}\,\Indicator{b=b'}.
\end{equation}
Furthermore, 
\begin{equation}\label{eqn:unique}
\mbox{for any $b,\ell$, there is at most one $a$ with $h_{a,b}^{(\ell)}\neq 0$.}
\end{equation}
These numbers can be constructed in $\OO(n_BL)$ time.
The $w^{(\ell)}$'s have $m^{O(1)}$-bit numerators and a common
$m^{O(1)}$-bit denominator.
\end{lemma}

\begin{observation}\label{obs:main}
Let $h_{a,b}^{(\ell)}$ and $w^{(\ell)}$ be as in Lemma~\ref{lem:algebraic}.
Define
\[ P^{(\ell)}[x,y]\ :=\ \sum_{z\in[n],\ a\in[n_A],\ b\in[n_B],\ \ell\in[L]} h_{a,b}^{(\ell)}A_a[x,z]B_b[z,y].
\]
Then the answer to Problem~\ref{prob:4matrix} is equal to
\[ Q\ :=\ \sum_{x,y\in[n],\ \ell\in [L]} w^{(\ell)} P^{(\ell)}[x,y]P^{(\ell)}[y,x].
\]
\end{observation}
\begin{proof}
By definition,
\[ Q\ =\ \sum_{x,y,z,z'\in[n],\ a,a'\in[n_A],\ b,b'\in[n_B],\ \ell\in[L]} w^{(\ell)}h_{a,b}^{(\ell)}h_{a',b'}^{(\ell)} A_a[x,z]B_b[z,y]A_{a'}[y,z']B_{b'}[z',x].
\]
By swapping $b$ and $b'$, and replacing $(x,z,y,z')$ with $(z,x,z',y)$,
\[ Q\ =\ \sum_{x,y,z,z'\in[n],\ a,a'\in[n_A],\ b,b'\in[n_B],\ \ell\in[L]} w^{(\ell)}h_{a,b'}^{(\ell)}h_{a',b}^{(\ell)} A_a[z,x]B_{b'}[x,z']A_{a'}[z',y]B_b[y,z].
\]
Because the input matrices are symmetric, summing yields:
\[ 2Q\ =\ \sum_{x,y,z,z'\in[n],\ a,a'\in[n_A],\ b,b'\in[n_B],\ \ell\in[L]} w^{(\ell)}(h_{a,b}^{(\ell)}h_{a',b'}^{(\ell)} +h_{a,b'}^{(\ell)}h_{a',b}^{(\ell)}) A_a[x,z]B_b[z,y]A_{a'}[y,z']B_{b'}[z',x].
\]
The observation then follows from (\ref{eqn:main}).
\end{proof}

With Observation~\ref{obs:main}, we can now easily reduce the problem to rectangular matrix multiplication:

\begin{theorem}\label{thm:main}
Let $n_A=2^{m/2}$, 
$n_B={m\choose k}$, and $L=n_A\cdot {m\choose \le k/2}$, where $m$ is divisible by $8$ and 
$k$ is even.
We can solve Problem~\ref{prob:4matrix} by computing
a product of an $nL\times nn_B$ matrix $\AAA$ with an
$nn_B\times n$ matrix $\BBB$,
and taking $\OO(n^2L)$ additional time.
Furthermore, the matrix $\AAA$ depends only on $A_0,\ldots,A_{n_A-1}$ and not
on $B_0,\ldots,B_{n_B-1}$.
\end{theorem}
\begin{proof}
For each $(x,\ell)\in [n]\times[L]$, $(z,b)\in [n]\times[n_B]$, and $y\in [n]$, define
\[ \AAA[(x,\ell),(z,b)]\ :=\ \sum_{a\in[n_A]} h_{a,b}^{(\ell)}A_a[x,z]\qquad
\mbox{and}\qquad \BBB[(z,b),y]\ :=\ B_b[z,y].
\]
Each entry of $\AAA$ can be computed in $\OO(1)$ time because of (\ref{eqn:unique}).

Compute the product $\AAA\BBB$.
Then $P^{(\ell)}[x,y] = (\AAA\BBB)[(x,\ell),y]$.  Knowing $P^{(\ell)}$,
we can compute $Q$ in $\OO(n^2L)$ additional time.
\end{proof}

\begin{corollary}
Suppose $n_A=n_B=n=2^{m/2}$ for some $m$ divisible by $8$. Suppose $t={m\choose k}$ and $L=n\cdot {m\choose \le k/2}$ for
some even $k$.
We can solve Problem~\ref{prob:4matrix} by computing
a product of an $nL\times nt$ matrix with an
$nt\times n^2/t$ matrix,
and taking $\OO(n^3L/t)$ additional time.
\end{corollary}
\begin{proof}
Apply Theorem~\ref{thm:main} to the $n$ matrices $A_0,\ldots,A_{n-1}$ and
a group $B_{jt,\ldots,(j+1)t-1}$ of $t$ matrices, for each $j\in [n/t]$.
(W.l.o.g., we may assume that $m$ is divisible by 8.)  This involves the product of an $nL\times nt$ and an $nt\times n$ matrix, for each $j$.
We return the sum of the $n/t$ output values.
Since the matrix $\AAA$ is common to all of the $n/t$ products,
these products can be combined into a single product, of an 
$nL\times nt$ and an $nt\times n^2/t$ matrix.
\end{proof}

The total running time above is
\[ \OO\left(   M\left(  n^2 {2\lg n \choose \le k/2},\  n{2\lg n\choose k},\  \frac{n^2}{{2\lg n\choose k}}
\right) \  +\ n^4 \frac{{2\lg n\choose \le k/2}}{{2\lg n\choose k}} \right).
\]
For example, setting $k\approx \frac{2}{75}\lg n$ yields
\[ \OO(M(n^{2.1156}, n^{1.2044}, n^{1.7957}) + n^{3.9113}).
\]
According to the latest bound~\cite{Uznanski26} on the rectangular
matrix multiplication exponents (which builds on OpenAI's $\omega\le 9/4$ result~\cite{MMOpenAI26}),
$\omega(a,b,c)\le \frac{a+b+c}{2} + \frac14 (ab/c+ac/b+bc/a)$
for $a\ge b\ge c > ab/(a+b)$.
Thus,  $\omega(2.1156,1.2044,1.7957)=\omega(2.1156,1.7957,1.2044)
< 3.957$.
Putting everything together, we obtain:

\begin{theorem}
The 3-uniform 4-hyperclique problem can be solved
in $O(n^{3.957})$ deterministic time.
\end{theorem}

Note that to ensure subquartic running time, we really need the latest rectangular matrix multiplication bounds (the pre-AI bounds are not good enough).  On the other hand, if $\omega=2$, the above bound would be
$O(n^{3.8})$ by setting $k\approx 0.08\lg n$.

By known reductions~\cite{LincolnWW18} (replacing $n$ with $2^{n/4}$), we immediately obtain the following consequence:

\begin{corollary}
Max-3-SAT for $n$ variables can be solved in $O(2^{0.99n})$ deterministic time. 
\end{corollary}



\subsection{Proof of the Algebraic Lemma}

It remains to prove Lemma~\ref{lem:algebraic}, which turns out to follow
from known techniques, in particular,
the work by Hrube\v{s}~\cite{Hrubes26} (whose motivation was on
a seemingly unrelated problem of Hurwitz).\footnote{
Curiously, 
Anthropic's ETH disproof~\cite{Antropic26} also mentioned Hurwitz's problem and cited 
\cite{Hrubes26}, though we have not studied their proof closely to see any connections.
}

\begin{lemma}\label{lem:poly}
For any even $k$, there is a degree-$(k/2)$ polynomial $p$ with $p(k)=1$ and
$p(i)=0$ for all even $i\in [k]$.
Furthermore, in $k^{O(1)}$ time, we can compute coefficients $c_0,\ldots,c_{k/2}\in\mathbb{Q}$ satisfying
$p(z) = \sum_{i=0}^{k/2} c_i {z\choose i}$.
\end{lemma}
\begin{proof}
We can take the polynomial $p(z)=\frac{1}{K}\prod_{i\in [k]\ \text{even}}(z-i)$, where $K=k(k-2)\cdots 2$.  The coefficients can be found by a standard inversion formula 
$c_i=\sum_{j=0}^i (-1)^{i-j}{i\choose j}p(j)$.
\end{proof}

The main tool is an anti-commuting family of  matrices:

\begin{lemma}[Anti-commuting family]\label{lem:anticommute}
Let $n_A=2^{m/2}$ for some $m$ divisible by $8$.
There exist $m$ $n_A\times n_A$ signed permutation matrices\footnote{
A signed permutation matrix has entries from $\{-1,0,1\}$, and each row or column has exactly one nonzero entry.
}
$E_0,\ldots,E_{m-1}$ that are orthonormal, anti-symmetric, and anti-commuting,
i.e., for each $a,a'\in [n_A]$,
\[  E_aE_a^T=I,\qquad E_a^T=-E_a,\qquad\mbox{and}\qquad  E_aE_{a'}=-E_{a'}E_a\ \ \mbox{if $a\neq a'$}.
\]
They can be constructed in $\OO(n_A)$ time.
\end{lemma}
\begin{proof}
This type of problem was first considered
by Hurwitz~\cite{Hurwitz1923} and Radon~\cite{Radon1922}. 
See \cite[Theorem~1]{GeramitaP74}, for example,
for a concise one-page recursive construction satisfying all these conditions.
\end{proof}

Let $\BBBB={[m]\choose k}$ and $\Lambda={[m]\choose \le k/2}$.
For any subset $\beta\in\BBBB$,
write $\beta=\{i_1,\ldots,i_k\}$ with $i_1<i_2<\cdots<i_k$, and 
define the signed permutation matrix $E_\beta=E_{i_1}\cdots E_{i_k}$.
Then for $k$ even,
\[ E_\beta E_\beta^T =  I\qquad\mbox{and}\qquad
E_\beta E_{\beta'} = (-1)^{|\beta\cap \beta'|} E_{\beta'}E_\beta\ \ \mbox{if}\ \beta\neq\beta'.
\]

For each $a\in [n_A]$, $b\in\BBBB$, and 
 $(j,\lambda)\in [n_A]\times\Lambda$, define
\[ h_{a,\beta}^{(j,\lambda)}\ :=\ E_\beta[a,j]\, \Indicator{\lambda\subseteq\beta}
\qquad\mbox{and}\qquad w^{(j,\lambda)} = c_{|\lambda|},
\]
where the $c_i$'s are from Lemma~\ref{lem:poly}.
Then
\begin{eqnarray*}
 \sum_{j\in [n_A],\ \lambda\in\Lambda}w^{(j,\lambda)} h_{a,\beta}^{(j,\lambda)}h_{a',\beta'}^{(j,\lambda)} &=& \sum_{j\in [n_A] ,\ \lambda\in\Lambda}
c_{|\lambda|} E_\beta[a,j]E_{\beta'}[a',j]\,\Indicator{\lambda\subseteq \beta\cap\beta'}\\
&=& \sum_{\lambda\in\Lambda}
c_{|\lambda|} (E_\beta E_{\beta'}^T)[a,a']\,\Indicator{\lambda\subseteq \beta\cap\beta'}\\
&=& p(|\beta\cap\beta'|)\cdot (E_\beta E_{\beta'}^T)[a,a'].
\end{eqnarray*}
Thus,
\begin{eqnarray*}
 \sum_{j\in [n_A],\ \lambda\in\Lambda}w^{(j,\lambda)} (h_{a,\beta}^{(j,\lambda)}h_{a',\beta'}^{(j,\lambda)}+h_{a,\beta'}^{(j,\lambda)}h_{a',\beta}^{(j,\lambda)}) 
&=& p(|\beta\cap\beta'|)\cdot (E_\beta E_{\beta'}^T + E_{\beta'}E_\beta^T)[a,a'].
\end{eqnarray*}
If $\beta=\beta'$, then this is $2\cdot \Indicator{a=a'}$.
Otherwise, if $|\beta\cap\beta'|$ is even, this is 0 because $p(|\beta\cap\beta'|)$ is zero;
if $|\beta\cap\beta'|$ is odd, this is also 0 because $E_\beta E_{\beta'}^T + E_{\beta'}E_\beta^T = 0$.
This proves property (\ref{eqn:main}),
when we equate $\BBBB$ with $[n_B]$, where $n_B={m\choose k}$, and
$[L]$ with $[n_A]\times \Lambda$, where $L=n_A\cdot {m\choose\le k/2}$.
Property (\ref{eqn:unique}) follows since $E_\beta$ is a signed permutation matrix.
\hfill\qed

\section{Second Approach}\label{sec:second}

The 3-uniform 4-hyperclique problem reduces to the following:
given four $3$-dimensional $N\times N\times N$ arrays $X,Y,Z,W$, compute 
\[ Q\ :=\ \sum_{a,b,c,d\in[N]}X_{a,b,c}Y_{a,b,d}Z_{a,c,d}W_{b,c,d}.
\]

W.l.o.g., assume that $N=2^n$ is a power of 2.
Embed the indices in $\bits$.
For $a\in\bits$, let $a_j$ denote the $j$-th bit of $a$, let
$\supp(a)$ be the set of all  indices $j$ for which the $j$-th bit of $a$ is 1,
and let $|a|=|\supp(a)|$.
Denote by $\oplus$ the bitwise-xor operation.

The key to the whole algorithm is the definition of the following cleverly concocted polynomial in one variable~$z$, with degree at most $n$ (and $n^{O(1)}$-bit integer coefficients):
\begin{align}
\Phi(z)\ :=&\ \sum_{a,b,c,d,t\in\bits:\ \supp(t)\subseteq \supp(c\oplus d)} 
z^{n-|c\oplus d|}(z-1)^{|t|} X_{a,b,c}Y_{a,b,d} Z_{a,c,d\oplus t} W_{b,c\oplus t,d}.
 \label{eq:weighted-conv}
\end{align}
The definition may look mysterious at first, but $\Phi(1)=Q$,
and so the answer to the problem can be obtained by summing all the coefficients of $\Phi$.

It turns out that $\Phi$ can be rewritten in a different way:

\begin{lemma}[Main identity] \label{lem:identity}
For each $a,b,e,\xi\in\bits$, let
\[ 
U_{a,b,e}(\xi)\ =\ \sum_{c\in\bits}(-1)^{\xi\cdot c}X_{a,b,c}Z_{a,c,c\oplus e}\qquad\text{and}\qquad
V_{a,b,e}(\xi)\ =\ \sum_{c\in\bits}(-1)^{\xi\cdot c}Y_{a,b,c\oplus e}W_{b,c,c\oplus e}
\]
Let $\ell(e,\xi):=n-|\supp(e)\cup\supp(\xi)|=|\{j\in[n]: (e_j,\xi_j)=(0,0)\}|$.  Then
\begin{equation}\label{eq:gen-fourier}
\Phi(z)\ =\ \frac{1}{2^n}\sum_{a,b,e,\xi\in\bits}(2z-1)^{\ell(e,\xi)}U_{a,b,e}(\xi) V_{a,b,e}(\xi).
\end{equation}
\end{lemma}

We defer the proof to Section~\ref{sec:identity}.  Roughly, both $U_{a,b,e}$ and
$V_{a,b,e}$ are Walsh--Hadamard transforms of certain functions (i.e., discrete Fourier transforms over $\bits$), and
(\ref{eq:gen-fourier}) is a Fourier inversion, with a certain polynomial
weight function in $z$.  The details in  Section~\ref{sec:identity} will reveal how
the definition of $\Phi$ comes about.

\subsection{Algorithm}

Assuming the above identity, we can design a subquartic algorithm to compute
all coefficients of $\Phi$.  We divide into two cases: low-degree coefficients
and high-degree coefficients.  We show that both cases reduce to rectangular matrix multiplication.
Intuitively, for the low-degree case, we use the original definition (\ref{eq:weighted-conv}):
low-degree terms here occur only when $|c\oplus d|$ is large, and there are not too many pairs $(c,d)$ to try.
On the other hand, for the high-degree case,
we use the identity (\ref{eq:gen-fourier}):
high-degree terms there occur only when
$\ell(e,\xi)$ is large, and there are not too many
pairs $(e,\xi)$ to try.

Let $\alpha_0\in (0,1)$ be a threshold parameter to be determined.

\paragraph{Low-degree coefficients.}\label{subsec:low-deg-coeff}
First, we describe how to compute the coefficient  of
$z^r$ in the polynomial $\Phi$, for a given $r\le \alpha_0 n$.
Here, we use the original definition (\ref{eq:weighted-conv}).

Fix $c,d\in\bits$ with $|c\oplus d|=h\ge n-r$.
Let $\lambda^{(c,d)}_i$ be the coefficient of $z^r$ in the polynomial $z^{n-h}(z-1)^i$ (if $h<n-r$, then $\lambda^{(c,d)}_i=0$).
For each $a,t,b\in\bits$ with $\supp(t)\subseteq\supp(c\oplus d)$ (there
are $2^h$ choices of $t$), let
\[ Z^{(c,d)}_{a,t} = \lambda^{(c,d)}_{|t|}Z_{a,c,d\oplus t}\qquad\mbox{and}\qquad
W^{(c,d)}_{t,b} = W_{b,c\oplus t,d}.
\]
Compute the product $Z^{(c,d)}W^{(c,d)}$ in $M(N,2^h,N)$ time.
Then the answer is equal to
$\sum_{c,d\in\bits:\ |c\oplus d|\ge n-r}\sum_{a,b\in\bits} X_{a,b,d} Y_{a,b,d} (Z^{(c,d)}W^{(c,d)})_{a,b}$, which requires $O(N^2 2^h)$ additional time.
Since there are $N\binom{n}{h}$ pairs $(c,d)$ at Hamming distance $h$, the running time is $\widetilde O\left(\sum_{h\ge (1-\alpha_0)n}N\binom{n}{h}M(N,2^h,N)\right)$.
Letting $h=(1-\alpha)n$ for $\alpha\le\alpha_0$, we can bound the expression by 
$\OO\left(\max_{\alpha\le \alpha_0} N\binom{\lg N}{\alpha \lg N} M(N,N^{1-\alpha},N)\right)$.

\paragraph{High-degree coefficients.}\label{subsec:high-deg-coeff}
Next, we describe how to compute the coefficients of $z^r$ in the polynomial $\Phi$, for a given $r > \alpha_0 n$.
Here, we switch to the identity (\ref{eq:gen-fourier}) from Lemma~\ref{lem:identity}.
Define $\Pi_k = \{(e,\xi): |\supp(e)\cup\supp(\xi)|=k\}$.
Then $|\Pi_k|=\binom{n}{k}3^k$ since there are $\binom{n}{k}$ ways to choose these $k$ coordinates of $\supp(e)\cup\supp(\xi)$, and at each such coordinate~$j$, we can choose $(e_j,\xi_j)$ from $\{(1,0),(0,1), (1,1)\}$.
Define $\Pi=\{(e,\xi): \ell(e,\xi)\ge r\}=\bigcup_{k=0}^{n-r}\Pi_k$.
Then $|\Pi|=\sum_{k=0}^{n-r}\binom{n}{k}3^k$.
Letting $k=(1-\alpha')n$ for $\alpha'\ge\alpha_0$,
we see that $|\Pi|\le n\cdot \max_{\alpha'\ge\alpha_0} \binom{n}{\alpha' n}3^{(1-\alpha')n}
=\OO\left(\max_{\alpha'\ge\alpha_0}  \binom{\lg N}{\alpha'\lg N} N^{(1-\alpha')\lg 3}\right)$.

We first compute $U_{a,b,e}(\xi)$ for all $a,b\in\bits$ and $(e,\xi)\in \Pi$. 
Fix $a\in\bits$. For each $b,c\in\bits$ and $(e,\xi)\in\Pi$, let
\[ X^{(a)}_{b,c} = X_{a,b,c}\qquad\mbox{and}\qquad Z^{(a)}_{c,(e,\xi)} = (-1)^{\xi\cdot c} Z_{a,c,c\oplus e}.
\]
It suffices to compute the product $X^{(a)}Z^{(a)}$ in $M(N,N,|\Pi|)$ time.
The total time over all $a\in\bits$ is $O(N\cdot M(N,N,|\Pi|))$.

Similarly, we compute $V_{a,b,e}(\xi)$ for all $a,b\in\bits$ and $(e,\xi)\in \Pi$.
Fix $b\in\bits$.  For each $a,d\in\bits$ and $(e,\xi)\in\Pi$ (think of $d=c\oplus e$), let
\[ Y^{(b)}_{a,d} = Y_{a,b,d}\qquad\mbox{and}\qquad 
W^{(b)}_{d, (e,\xi)} = (-1)^{\xi\cdot (d\oplus e)} W_{b,d\oplus e,d}.
\]
It suffices to compute the product $Y^{(b)}W^{(b)}$ in $M(N,N,|\Pi|)$ time.
The total time over all $b\in\bits$ is $O(N\cdot M(N,N,|\Pi|))$.

Let $\lambda_i$ be the coefficient of $z^r$ in $(2z-1)^i$. 
The answer is $\frac{1}{2^n}\sum_{a,b\in\bits,\ (e,\xi)\in\Pi}\lambda_{\ell(e,\xi)} U_{a,b,e}(\xi)V_{a,b,e}(\xi)$, which requires $O(N^2|\Pi|)$ additional time.
(If $(e,\xi)\not\in\Pi$, then $\lambda_{\ell(e,\xi)}=0$.)

\paragraph{Total running time.}
The overall time bound for a fixed $\alpha_0$ is bounded by
\[ \OO\left(\max_{\alpha\le\alpha_0\le\alpha'} \left(N\cdot \binom{\lg N}{\alpha \lg N} M(N,N^{1-\alpha},N)\ +\ N\cdot M\left(N,N,\binom{\lg N}{\alpha'\lg N} N^{(1-\alpha')\lg 3}\right)\right)\right).
\]
Using the latest bound~\cite{Uznanski26} on the rectangular
matrix multiplication exponents
$\omega(1,c,1)=\omega(1,1,c)\le 1+c+1/(4c)$,
and setting $\alpha_0=0.455$ (and noting the maximum occurs when $\alpha=\alpha'=\alpha_0$), we obtain:

\begin{theorem}
The 3-uniform 4-hyperclique problem for $N$ vertices can be solved
in $O(N^{3.998})$ deterministic time.
\end{theorem}

Note that to ensure subquartic running time, we also need the latest rectangular matrix multiplication bounds here.  If $\omega=2$, the above bound would be
$O(N^{3.95})$ (which is worse than in Section~\ref{sec:first}).



\subsection{Proof of the Algebraic Identity}\label{sec:identity}

It remains to prove Lemma~\ref{lem:identity}.  This is where we will try to demystify
the definition of $\Phi$ and its identity.

For each $a,b,e\in\bits$ and $c\in\bits$,
define 
\[ f_{a,b,e}(c)\ :=\ X_{a,b,c}Z_{a,c,c\oplus e}\qquad\mbox{and}\qquad
g_{a,b,e}(c)\ :=\ Y_{a,b,c\oplus e}W_{b,c,c\oplus e}.
\]
Then $U=\widehat{f_{a,b,e}}$ and $V=\widehat{g_{a,b,e}}$ are just the
Walsh--Hadamard transforms\footnote{
Given $f:\bits\rightarrow\bits$,
its Walsh--Hadamard transform $\widehat{f}:\bits\rightarrow\bits$ is defined as
$\widehat{f}(\xi)=\sum_{c\in\bits} (-1)^{\xi\cdot c} f(c)$.
} of $f_{a,b,e}$ and $g_{a,b,e}$.
For each $s\in\bits$, define the ``weight function''
\begin{equation*}
K_{z,e}(s)\ :=\ 
\begin{cases}
z^{n-|e|-|s|}(z-1)^{|s|} & \text{if }\supp(s)\cap \supp(e)=\emptyset\\
0 &\text{otherwise.}
\end{cases}
\end{equation*}
By a change of variable $d=c'\oplus e$ and $t=c\oplus c'$,  the definition
of $\Phi$ can be rewritten as
\begin{eqnarray*}
\Phi(z) &=& \sum_{a,b,c,c',e\in\bits:\ \supp(c\oplus c')\cap\supp(e)=\emptyset} 
z^{n-|c\oplus c'\oplus e|}(z-1)^{|c\oplus c'|} X_{a,b,c}Y_{a,b,c'\oplus e} Z_{a,c,c\oplus e} W_{b,c',c'\oplus e}\\
&=& \sum_{a,b,c,c',e\in\bits}
K_{z,e}(c\oplus c') f_{a,b,e}(c)g_{a,b,e}(c')\\
&=& \sum_{a,b,e\in\bits} (K_{z,e}\star f_{a,b,e}\star g_{a,b,e})(0),
\end{eqnarray*}
where $\star$ denotes convolution.
By Fourier inversion\footnote{
$f(c)=\frac{1}{2^n}\sum_{\xi\in\bits} (-1)^{c\cdot\xi}\widehat{f}(\xi)$.
} 
evaluated at 0,
\[ \Phi(z) \ =\ \frac{1}{2^n}\sum_{a,b,e,\xi\in\bits}\reallywidehat{K_{z,e}\star f_{a,b,e}\star g_{a,b,e}}(\xi)\ =\ 
\frac{1}{2^n}\sum_{a,b,e,\xi\in\bits}\widehat{K_{z,e}}(\xi)U_{a,b,e}(\xi)V_{a,b.e}(\xi).
\]
It remains to evaluate $\widehat{K_{z,e}}(\xi)$. 
Write $K_{z,e}(s)=\prod_{j=1}^nk_{z,e_j}(s_j)$, where 
we define $k_{z,0}(0)=z$, $k_{z,0}(1)=z-1$, $k_{z,1}(0)=1$, and $k_{z,1}(1)=0$. 
Now, $\widehat{k_{z,e_j}}(\xi_j) = \sum_{s_j\in\{0,1\}} (-1)^{\xi_js_j}k_{z,e_j}(s_j) = k_{z,e_j}(0)+ (-1)^{\xi_j}k_{z,e_j}(1)$
is equal to 
\[\left\{\begin{array}{ll}
z+(z-1)=2z-1 &\mbox{if $(e_j,\xi_j)=(0,0)$}\\
z-(z-1)=1 &\mbox{if $(e_j,\xi_j)=(0,1)$}\\
1+0=1 &\mbox{if $(e_j,\xi_j)=(1,0)$}\\
1-0=1 &\mbox{if $(e_j,\xi_j)=(1,1)$.}    
\end{array}\right.\]
Therefore, $\widehat{K_{z,e}}(\xi)=\prod_{j=1}^n \widehat{k_{z,e_j}}(\xi_j) = 
\prod_{j:\, e_j=\xi_j=0}(2z-1)=(2z-1)^{\ell(e,\xi)}$,
as desired.
\hfill\qed

\section*{AI Disclosure}

Both proofs in this paper are obtained using ChatGPT Sol 6.1,
prompted by different suggestions.
This manuscript is human-written: we (the authors) rewrote the proofs and presented them in our own ways, aiming for clear and broadly accessible exposition
(the initial solutions were not easy to understand).
The authors verified the correctness and originality of all content, including references.

\small
\bibliographystyle{alphaurl}
\bibliography{ref}

\end{document}